\documentclass{ws-procs961x669}            % default, citations in superscript

\newcommand{\ba}{\begin{eqnarray}}
\newcommand{\ea}{\end{eqnarray}}
\newcommand{\beqs}{\begin{eqnarray}}
\newcommand{\eeqs}{\end{eqnarray}}

\begin{document}
\title{Generelized parton distributions and the structure of the hadrons}
%Using World Scientific's WS-procs961x669\\
%document class in \LaTeX2e
% }

\author{O.V. Selyugin}

\address{BLTPh, Joint Institut for Nuclear Research, \\
  Dubna, 141986, Russia\\
%University Department, University Name,\\
%City, State ZIP/Zone, Country\\
$^*$E-mail: ab\_selugin@theor.jinr.ru\\
}

\begin{abstract}
   The dependence  of the hadron interaction on  its structure is examined in the framework
   of the generelized parton distributions (GPDs). The $x$ dependence of the GPDs is determined
   by the parton distribution functions (PDFs), which were obtained from the deep inelastic scattering.
   The analysis of the whole sets of experimental data on the electromagnetic form factors of the proton and neutron
   with taking into account  many forms of PDFs, obtained by the different Collaborations, make it possible  to obtain
   the  special momentum transfer dependence of the GPDs. This permits us to obtain the electromagnetic and
   gravitomagnetic form factors of the nucleons. The impact parameter dependence of  the proton and neutron charge and matter densities is examined. The elastic hadron scattering at high energies
   was analyzed in the framework of the model that takes into account both these form factors
   (electromagnetic and gravitomagnetic).
\end{abstract}


\bodymatter

\section{Introduction}

     The hadron diffraction processes at high energies give the valuable information about hadron structure.
    Especially it concerns the processes examined at the LHC with energy from $\sqrt{s} = 7$ TeV
    up to  $\sqrt{s} = 14$ TeV \cite{Rev-LHC}.
    At these high energies the exchange of the mass Reggions with intercept less than unity
    can be neglected and the diffractive scattering is determined
     in most part by the cross-even massless Region  - Pomeron
    and, possibly, the cross-odd Reggion -Odderon.
    The optic theorem links the imaginary part of the elastic scattering
    amplitude with the size of the total cross sections.
    The size and energy dependence of the total cross sections
    is connected intimately with the basic principle of quantum field theory and
    it is determined by the basic structure  of the interaction  hadrons,
    especially its radius of the interaction which is tightly connected with the hadron form factors.
    The first new data on the differential cross sections obtained at LHC
    do not coincide with any model predictions. This requires essential modernization of the phenomenological models.
    However, the new data on the total cross section, obtained by the two different Collaborations (TOTEM-CMS and group of ATLAS),
    again (as in the case of Tevatron) show a large difference which  reaches up to $6$ mb at $\sqrt{s} = 8 $ TeV.
    The sizes of the total cross sections are obtained from the analysis of the differential cross sections  obtained at small
    momentum transfer with some model assumptions about the behavior of the hadron elastic scattering amplitude.

    The hadron structure is represented in the structure of the scattering amplitude.
    It has to satisfy the basic principles of  quantum field theory - casuality, analyticity, et cetera.
     The structure of the elastic scattering amplitude
     is represented by the  imaginary and real parts and its energy and momentum transfer dependence.
      The basic properties require that the scattering amplitude should
    satisfy the integral dispersion relations.   They tie the sizes and
     energy dependence of the real and imaginary parts of the elastic scattering amplitude.
     Hence, it is very important to determine the form and energy dependence of the real part
     and the ratio of the real to imaginary parts of the scattering amplitude - the value
     $\rho(s,t)$. To our regret the first experimental data obtained at the LHC at $\sqrt{s} = 7$ TeV
     reveal  some problems in the determination of the total cross sections and the size
     of $\rho(s= 7)$  TeV, $t=0)$ \cite{Sel-NPA14}.
     The next experiment at $\sqrt{s}=8$ TeV
     shows that the nonlinear behavior of the differential cross sections at small momentum transfer
     exist at such large energy too.
     Such non-linear behavior, of course, is connected  with hadron structure.
     In most part it can be determined by the contributions of  the meson cloud surrounding the hadron.
%     \cite{76,97}.
       It leads to some problems in the determination of the total cross section
     and parameter $\rho(s,t=0)$ \cite{Sel-NPA14}.

     The form factors can represent either the electromagnetic structure of the hadron or the
     matter distribution into the hadron.
              The different reactions   can be related with the different form factors.
  It is very likely that the strong hadron-hadron interactions can be proportional to the matter distribution of the hadrons.

     Both form factors can be obtained from the one function - Generelized parton distributions.
     The remarkable properties of the $GPDs$ is that the integration of  different momentum of
     GPDs over $x$ gives us different hadron form factors  \cite{Mil94,Ji97,R97}.

    The $x$ dependence of $GPDs$ in most part is determined by the standard
    PDFs which are obtained by the different Collaborations from the analysis of
    the dip-inelastic processes.
     We can write the factorization form of GPDs - ${\cal{F}}_{i} = PDFs_{i}(x) f(x,t)$ \cite{ST-PRDGPD}.
      The main part of the $x$ dependence of $GPDs$ is reflected in the standard parton distribution functions (PDFs)
      which are obtained from the analysis of the deep inelastic reactions by many different collaborations
     using the different forms of the $x$ dependence of the PDFs.
     However, to obtain the true form of the form factors,  it is needed to know the true form of the function $f(x,t)$
     which  contains both the $x$  and the $t$ dependence.

 \section{Hadron form factors}

    On the basic of our analysis of  many forms of the PDFs and  practically all available experimental
       data on the electromagnetic form factors of the proton and neutron
       (the isotopic invariance can be used to relate the proton and neutron GPDs)
       % .we are determining the function $f(x,t)$
     the most preferable form of the PDFs   was chosen \cite{GPD-PRD14}.

  Taking  $f(x)$ in the form
   $ f_{u}(x) =  [(1-x)^{2+\epsilon_{u}}]/[(x_{0}+x)^{m}]$ and $f_{d}(x) = (1+\epsilon_{0}) [(1-x)^{1+\epsilon_{d}}]/[(x_{0}+x)^{m}]$
   we can obtain different parts of GPDs
\begin{eqnarray}
{\cal{H}}^{u} (x,t) \  && = q(x)^{u}  \   e^{2 a_{H}   f(x)_{u}  \ t };  \ \ \   \\ \nonumber
{\cal{H_d}}^{d} (x,t) \ &&  = q(x)^{d}  \   e^{2 a_{H} f_{d}(x)  \ t };
\label{t-GPDs-H}
\end{eqnarray}
\begin{eqnarray}
%\ba
{\cal{E}}^{u} (x,t) \ && = q(x)^{u} (1-x)^{\gamma_{u}} \   e^{2 a_{E}  \  f(x)_{u}  \ t }; \ \ \   \\ \nonumber
{\cal{E_d}}^{d} (x,t) \ && = q(x)^{d}  (1-x)^{\gamma_{d}} \   e^{2 a_{E} f_{d}(x) \ t },
\label{t-GPDs-E}
%\ea
\end{eqnarray}

   The hadron form factors were obtained     by the numerical integration over $x$
   and then
    by fitting these integral results for the electromagnetic form factors
    by the standard dipole form with some additional parameters.
    It allows us to find  separate contributions of the $u$ and $d$ quarks to the different parts of
   the electromagnetic
    form factors \cite{GPD-PRD14}. % ( see Fig.1 a,b).

  The integration of the next moment of the GPDs gives us the mattter form factors,
    which was fitted by the also standard dypol form. It gives the size of the
    $\Lambda^{2}_{grav.}=1.6$ GeV$^2$.

\section{Hadron form factors and elastic nucleon-nucleon scattering}

      Note that the transition form factor $G^{*}_{M}(t)$
      for the magnetic $N \rightarrow \Delta $ is depended only
      from the difference of $ E_{u}(x,\xi=0,t) $ and $ E_{d}(x,\xi=0,t) $.
      It is gives the unique possibility to check the momentum transfer dependence of
      spin depended part of GPDs.
The relevant $GPD_{N\Delta}$ can be expressed in terms of the isovector GPD,
 taking into account $k_{v}=k_{p} - k_{n} =3.70 $ ,
    yielding the sum rule \cite{Guidal}
%\ba
\begin{eqnarray}
  G^{*}_{M} (t) = \frac{G^{*}_{M} (t=0)}{k_{v} } \int_{-1}^{1} dx ( E_{u}(x,\xi,t) - E_{d}(x,\xi,t) ).
\end{eqnarray}

   Figure 1a
   shows a sufficiently good coincidence with experimental data.
   It is confirmed that the form of the momentum transfer dependence of the $E(x,\xi,t)$ determined in our model  is right.

   It allows us to use
  both hadron form factors ( electromagnetic and gravitomagnetic) % were used in the framework of
  to build  a high energy generalized structure  model  of the elastic nucleon-nucleon scattering
  % That allow us to build the model
  with minimum fitting  parameters \cite{HEGS0,HEGS1,NP-hP}.

   The Born term of the elastic hadron amplitude can now be written as
   \begin{eqnarray}%\ba
 F_{h}^{Born}(s,t)= F_{IP}(\epsilon,s,t) \ (1+\frac{r_1}{\hat{s}^{0.5}})
 %   \\ \nonumber
     +  F_{3g}^{even}(\epsilon,s,t) + F_{3g}^{odd}(\epsilon,s,t) (1+\frac{r_2}{\hat{s}^{0.5}}),
    \label{FB}
% \nonumber
\end{eqnarray}
 with $   \hat{s}=s \ e^{-i \pi/2}/s_{0}$ ;  $s_{0}=4 m_{p}^{2} \ {\rm GeV^2}$.
  The intercept $1+\epsilon_{1} =1.11$ was chosen from the data of different reactions
  and was fixed by the same size for all terms of the Born  scattering amplitude.
 The slope of the scattering amplitude has the standard logarithmic dependence on the energy
% \begin{eqnarray}%\ba
 $   B(s) = \alpha^{\prime} \ ln(\hat{s}) $
% \nonumber
%\end{eqnarray}
  with $\alpha^{\prime}=0.24$ GeV$^{-2}$  and has some small additional term \cite{HEGS1},
    which reflects the small non-linear behavior of  $\alpha^{\prime}$ at small momentum transfer \cite{Sel-Df16}.
The final elastic  hadron scattering amplitude is obtained after unitarization of the  Born term.
    So, at first, we have to calculate the eikonal phase
%   \begin{eqnarray}
 $ \chi(s,b) \   =  -\frac{1}{2 \pi}
   \ \int \ d^2 q \ e^{i \vec{b} \cdot \vec{q} } \  F^{\rm Born}_{h}(s,q^2)  $,
% \label{tot02}
% \end{eqnarray}
  and then obtain the final hadron scattering amplitude
    \begin{eqnarray}
 F_{h}(s,t) = i s
    \ \int \ b \ J_{0}(b q)  \ \Gamma(s,b)   \ d b\, ; \ \  \ \ \ \
% \label{overlap}
% \end{eqnarray}
%     \begin{eqnarray}
   {\rm with }  \ \ \ \   \Gamma(s,b)  = 1- \exp[ \chi(s,b)] .
% \label{overlap}
\end{eqnarray}
     At large $t$  our model calculations are extended up to $-t=15 $ GeV$^2$.
  We added a small contribution of the energy  independent  part
  of the spin flip amplitude in the form similar to the proposed    in \cite{Kuraev-SF} and analyzed in \cite{W-Kur}.
 % \begin{eqnarray}%\ba
 % $ F_{sf}(s,t) \ =  h_{sf} q^3 F_{1}^{2}(t) e^{-B_{sf} q^{2}}$.
% \nonumber
 % \end{eqnarray}
  The model is very simple from the viewpoint of the number of fitting parameters and functions.
  There are no any artificial functions or any cuts which bound the separate
  parts of the amplitude by some region of momentum transfer.
        In the framework of the model the description of the experimental data was obtained simultaneously
        at the large momentum transfer and in the Coulomb-hadron region in the energy region from $\sqrt{s}=9 $ GeV
        up to LHC energies \cite{NP-Dmin}. Figure 1b presents  % the description at  $\sqrt{s}=11.4 $ GeV and Figure 3(b)
        the model predictions for $\sqrt{s}=13 $ TeV, that are in  very good coincidence with the preliminary data
        presented at the conference BLOIS-17 (2017) \cite{D-Blois17}.

%%%Fig 1
\begin{figure}
\begin{center}
\includegraphics[width=.45\textwidth]{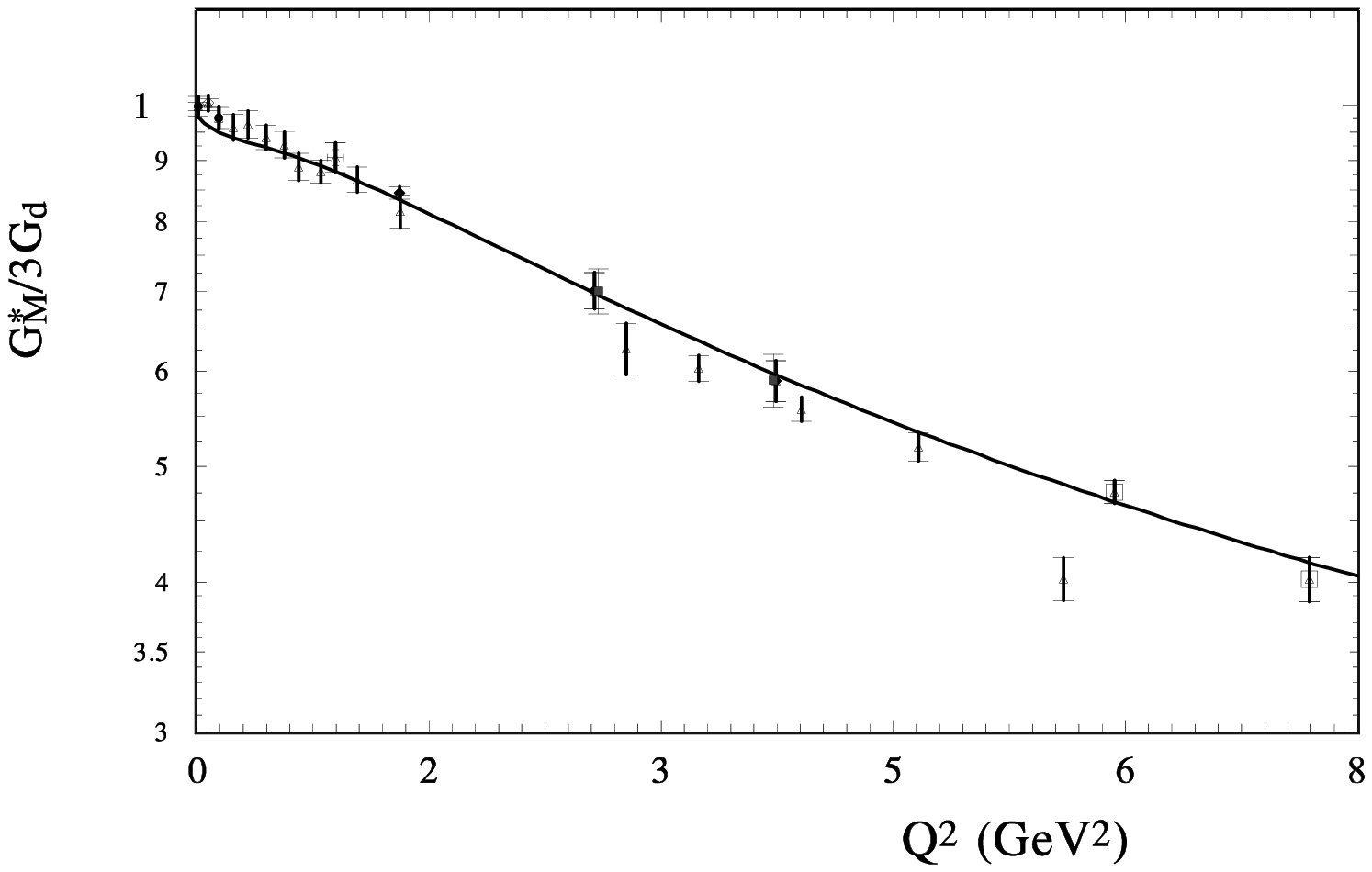} %{ffdm.ps}
\includegraphics[width=.45\textwidth]{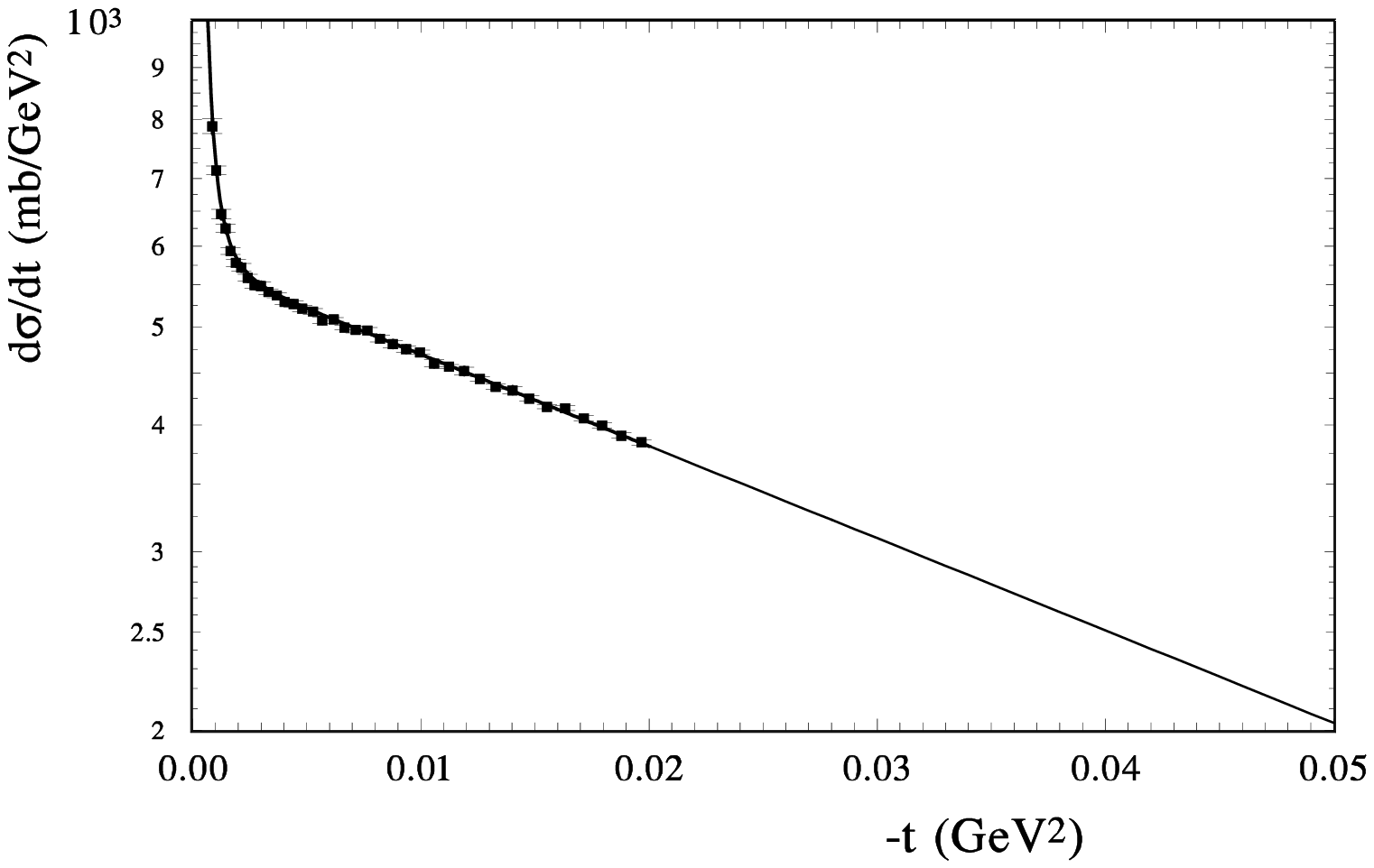} %{ffdm.ps}
\caption{a) [left]
 a) [left] Transition form factor $G^{*}_{M}(t)$ (line -the calculations in the framework of the model,
           points - the experimental data \cite{G-data};
b) [right]  Differential cross sections of the elastic $pp$ scattering are predicted
  by the HEGS model (line) and preliminary data of the TOTEM Collaboration.
% and data of the TOTEM   Collaboration
   \cite{D-Blois17} % \cite{T737,Blois17} )
  }
 \end{center}
\label{Fig1}
\end{figure}

%%%Fig 2
\begin{figure}
\begin{center}
\includegraphics[width=.45\textwidth]{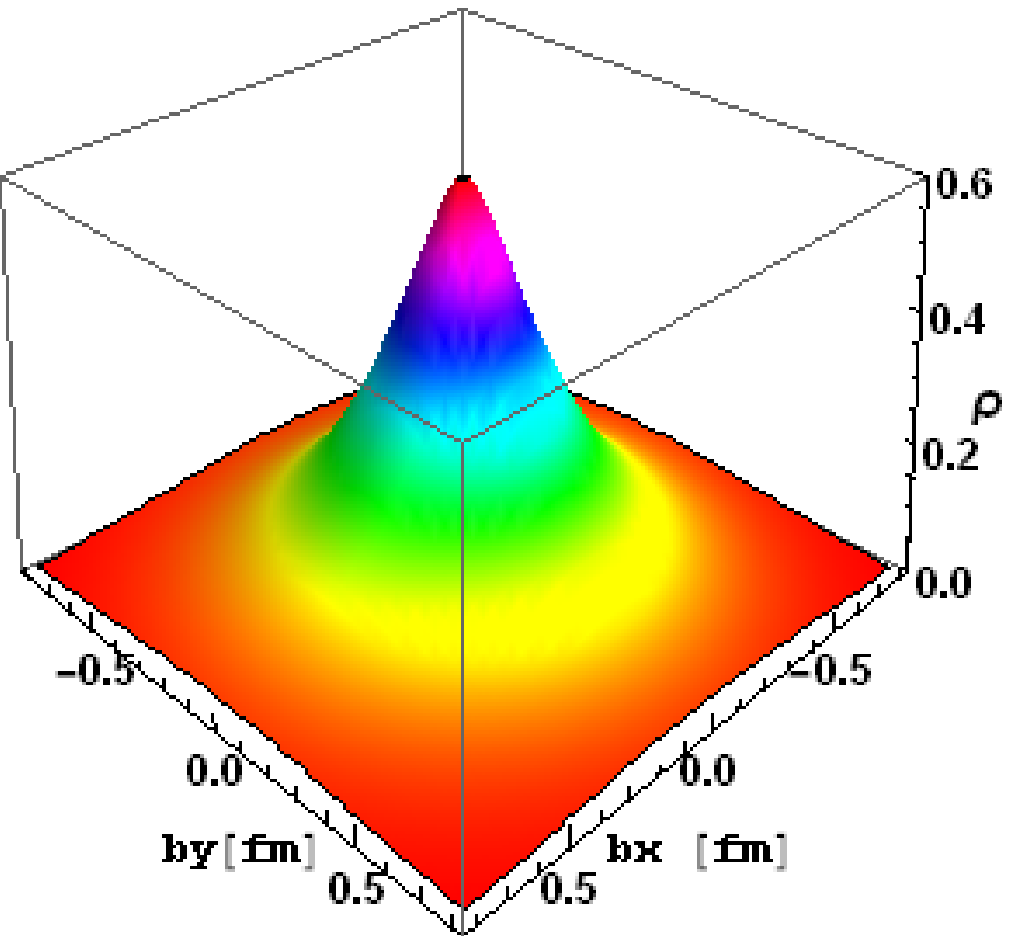} %{ffdm.ps}
\includegraphics[width=.45\textwidth]{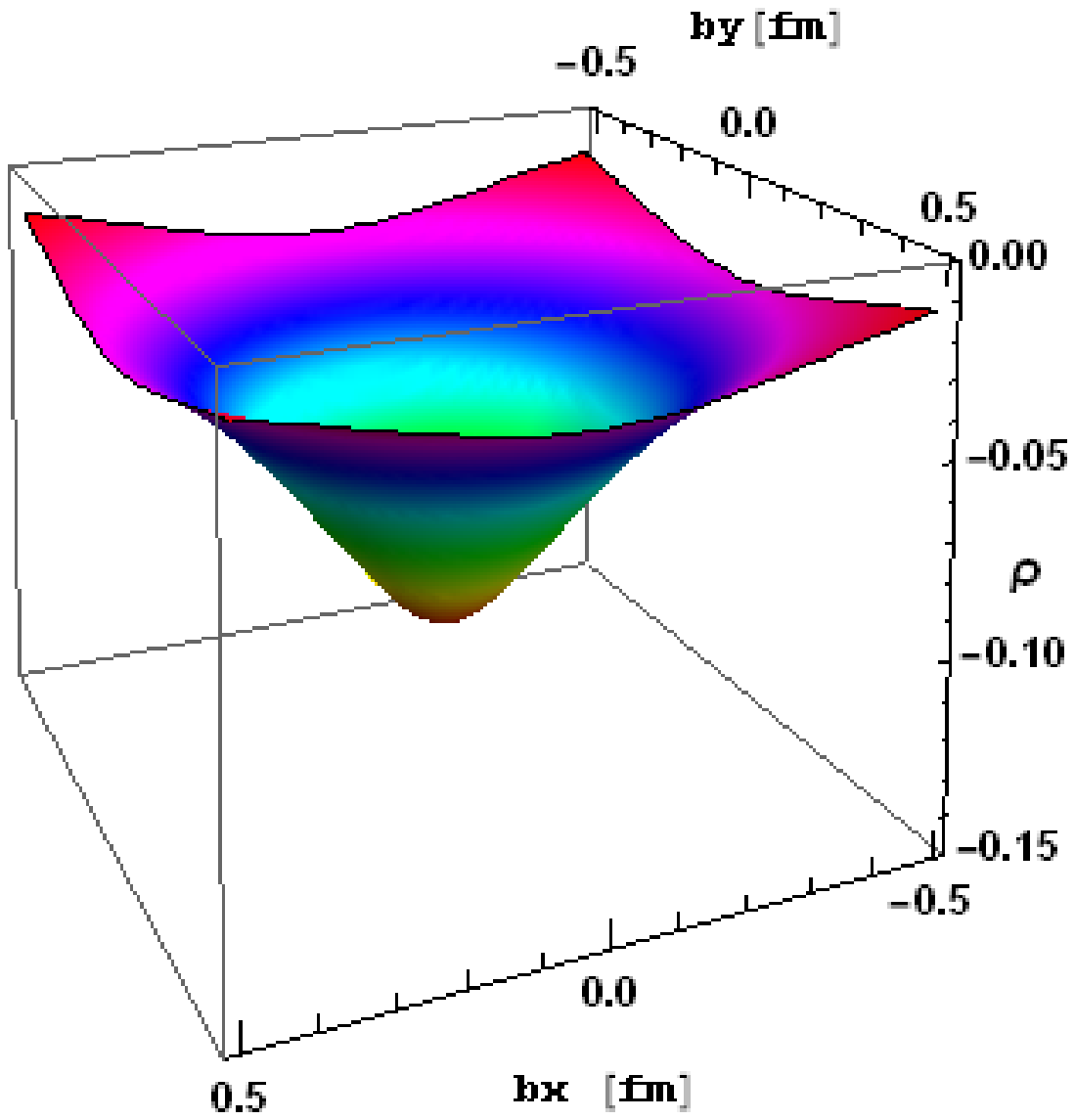} %{ffdm.ps}
\caption{a) [left] Charge density of  $u$-quark
 b) [right]  charge density of  $d$-quark .
}
\end{center}
\label{Fig2_}
\end{figure}

\section{Structure of nucleons in the impact parameter representation}

  The hadron structure is most obviously presented in the impact parameter  % \cite{BurkT,SopperT,RayT}
  representation  \cite{BurkT,RayT}.
  The charge density can be obtained by the integration of the GPDs over $x$ and $t$.
  \begin{eqnarray}
 \rho_{G_{E}}(b)
% & = & \int d^{2}q [F_1 (q^2)+ \tau F_{2}(q^2)] \ e^{i \vec{q} \vec{b} } \nonumber \\
  =  \sum_{q} e_{q} \int^{1}_{0}  dx \int d^{2}q [ H_{q}(x,\xi=0,q^2)
 +  \tau E_{q}(x,\xi=0,q^2)] e^{i \vec{q} \vec{b} } % & =& \int_{0}^{\infty} \ \frac{q dq}{2 \pi} J_{0}(q b) \
% G_{E}(q^2).
\end{eqnarray}
 where $\tau = (q / m_{p})^2 $.
 In the framework of our model of $t$-dependence of  GPDs the calculation of
  both forms of the hadron charge distribution in the impact parameter representation
  can be carried out
  and, moreover,  we can determine
   the separate  contributions of $u$ and $d$ quarks (see Fig.2).
% The respective
%  separate contributions of $u$ and $d$-quarks are shown in Fig. 4 (a).
It gives the possibility to compare the distribution of the electric charge and matter (that is, gravitational charge) in the nucleon
%with the matter density,
 which was  obtained from  the integration of the next momentum of GPDs.
%\begin{eqnarray}
% \rho_{0}^{Gr}(b) = \frac{1}{2 \pi}
%\int^{0}_{\infty} \  dq \ q  \ J_{0}(q b) A(q^2).
%\end{eqnarray}
 The difference of the charge density and the matter density of $u$ quark  and $d$ quark %(c.f. Section 5)
 for proton is shown in Fig.3(a,b).

%%%Fig 3
\begin{figure}
\begin{center}
\includegraphics[width=.45\textwidth]{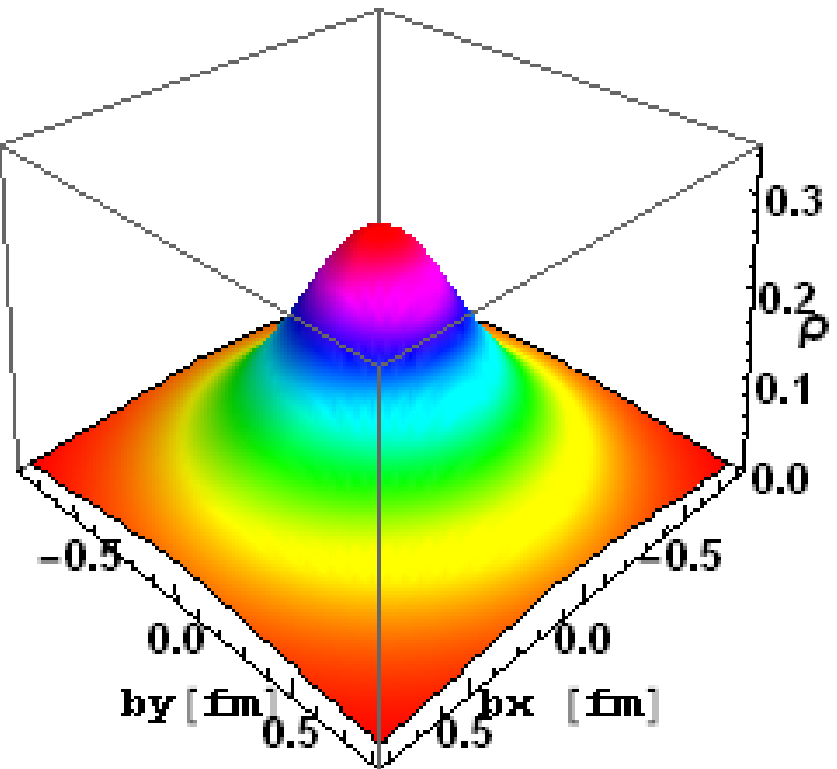} %{ffdm.ps}
\includegraphics[width=.45\textwidth]{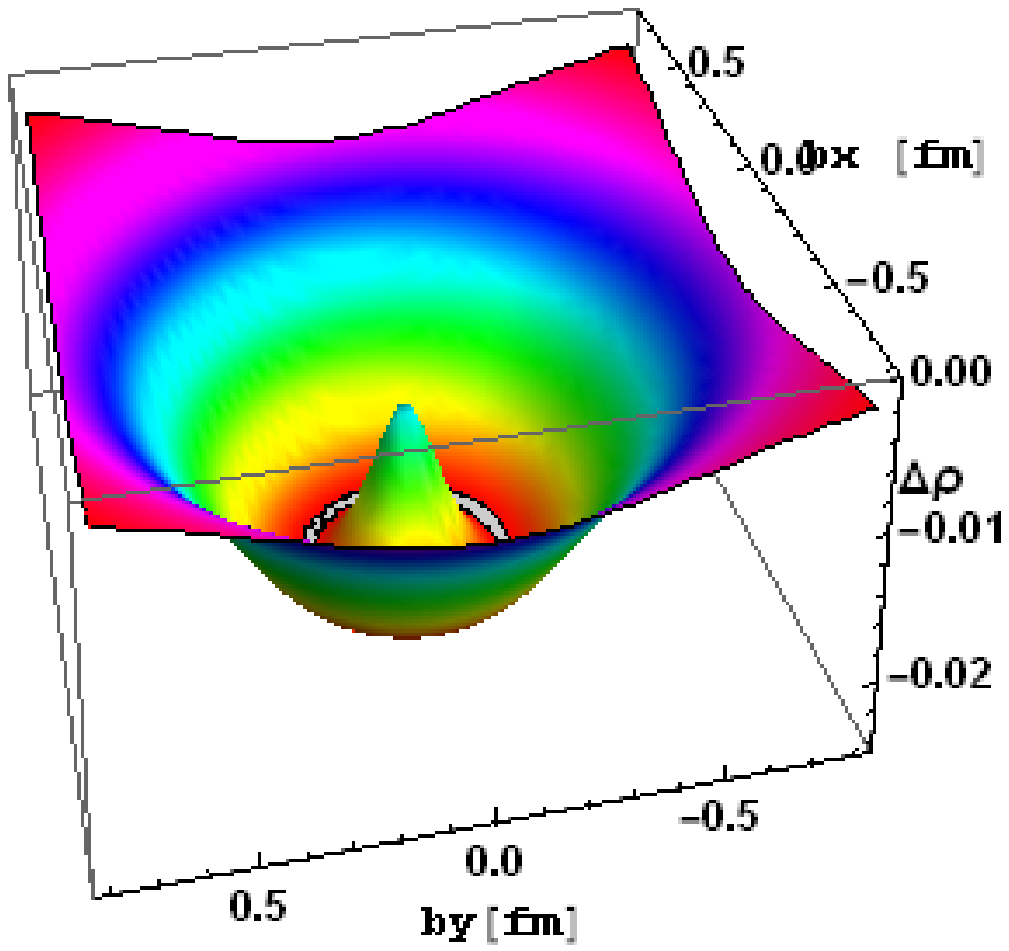} %{ffdm.ps}
\caption{a) [left] Difference between the charge density and matter dencity of $u$-quark;
[right] difference between the charge density and matter density for $d$-quark;
}
\end{center}
\label{Fig3_}
\end{figure}

\section{Conclusion}

    The generalized parton distributions (GPDs) represent the basic properties of the hadron
    structure.
  The different properties of  the hadron structure are reflected in the elastic scattering amplitude.
  One main part of it is the corresponding form factors of the nucleons.
  Our form of the momentum transfer dependence of the GPDs allow us to
  calculate the electromagnetic form factors, gravitomagnetic form factors, transition form factors,
 and Compton form factors.
As a result, the description of various reactions is based on the same
representation of the hadron structure.
The results give good coincidence with the available  experimental data.
                      Especially it concerns the  real Compton scattering and
                     high energy elastic hadron scattering.
  The new high energy generalized structure (HEGS) model, based on the electromagnetic and gravitomagnetic form factors,
   gives a good quantitative description of the existing experimental data
  of the proton-proton and proton-antiproton elastic scattering
  in a wide region of the energy scattering and momentum transfer, including the Coulomb-hadron interference region,
  with the minimum number of fitting parameters.
     Its predictions describe  quantitatively  the  experimental data obtained at the LHC at $\sqrt{s}=7$ and $\sqrt{s}=8$ TeV,
  and are in good agreement  with recent preliminary data at $\sqrt{s}=13$ TeV \cite{D-Blois17} taking into account only
  statistical errors but with some small additional normalization coefficient.
    In the impact parameter representation the GPDs give us a separate contribution of the
    density of charge and matter distributions of the quarks.

\vspace{0.5cm}
{\bf Acknowledgments}
%\section{Acknowledgments}
 {\it The authors would like to thank  V. Petrov for the invitation
   to take apart in the conference and J.-R. Cudell
   for fruitful   discussion of some questions   considered in the paper.}

%\section*{References}

\end{document}